# Nimble@ITCEcnoGrid: A Grid in Research Domain for Weather Forecasting

Vijay Dhir[1], Dr. Rattan K Datta[2] and Dr. Maitreyee Dutta[3]

[1] Assistant Professor, Department of Information Technology, SBBSIET, Punjab, India.
`vijaykdhir@yahoo.com`
[2] Retd. Advisor Govt. of India, MET, New Delhi, India
`rkdatta_in@yahoo.com`
[3] Associate Professor, Deptt. of CSE, NITTTR, Chandigarh, India
`d_maitreyee@yahoo.co.in`

## ABSTRACT

*Computer Technology has Revolutionized Science. This has motivated scientists to develop mathematical model to simulate salient features of Physical universe. These models can approximate reality at many levels of scale such as atomic nucleus, Earth's biosphere & weather/climate assessment. To solve these type of complex problems in usable time frame, there is a need of high performance powerful computer mechanism which can do the calculations in a time bound with high precision. If the computer power is greater, the greater will be the accuracy in approximation i.e. close will be the approximation to the reality. The speed of the computer required for solution of such problems require computers with processing power of teraflops to Pets flops speed.. The way to speed up the computation is to "parallelize" it i.e. divide the work into modules that can be worked on by separate processors at the same time. Thus we can solve the problems that are Non-Polynomial form in polynomial time.*

*One of the approach is to use multimillion dollar Supercomputer or use Computational Grid ( Which is also called poor man's supercomputer) having geographically distributed resources e.g. SETI@home (Used to detect radio waves emitted by intelligent civilizations outside earth) has 4.6 million participants computers. There are many alternatives tools available to achieve this goal like Globus Toolkit, Entropia, Legion, BOINC etc but they are mainly based on Linux platform. As majority of the computers available are windows based, so it will be easy to develop a larger network of computers which will use the free cycles of the computer to solve the complex problem at window platform.*

*Nimble@ITCEcnoGrid has been developed. It includes the feature of Inter Thread Communication which is missing in any of the toolkits available. Nimble@ITCEcnoGrid Framework (A Fast Grid with Inter-thread communication with Economic Based Policy) was tested for computation of 'PI' up to 120 decimal points. Encouraged by the speed the same system has been utilized to computes the Momentum, Thermodynamics and Continuity equations for the Weather Forecasting using the Windows based Desktop computers..*

## 1. INTRODUCTION

According to Moore's law **[1],** the speed of chip doubles in every 18 months it is because of Moore's law Pc's become more fast and there are millions of fast computer connected through Internet. The idea of using these computers as grid.

The Foster-Kesselman duo organized in 1997, at Argonne National Laboratory, a workshop entitled "Building a Computational Grid" .At this moment the term "Grid" was born. The workshop was followed in 1998 by the publication of the book "The Grid: Blueprint for a New Computing Infrastructure" by Foster and Kesselman **[2]** themselves. For these reasons they are considered the fathers of the Grid.

Grid computing very promising many independent computer are grouped together to solve a particular complex problem. According to Foster-Kesselman **[3]** " Grid Computing is a special type of parallel Computing which relies on complete computers (with onboard CPU, Power supply, network interface etc) connected to the internet by conventional network interface, such as Ethernet".

So, instead of investing so much of money on Super Computer, we can built economical Computational grid having the power same that of the desktop computer. SETI@home **[4]** has been very successful in this regard. It has 4.6 million participants, out of which 600,000 remain active.

## 2. LITERATURE REVIEW OF DESKTOP GRID

A brief description of various tools is given below and compared with Nimble..
Alchemi**[5]** is the .NET framework that provides runtime machinery and programming environment requires constructing desktop Grid. It supports object oriented programming in addition to file based job model cross platform is provided by web server interface. It has no Inter thread communication.

Entropia **[6]** uses a window desktop grid system by aggregating the raw desktop resources into a single logical resource. There is a one centralized computer, which administrates various desktop clients. But it does not provide web server interface for cross platform.

SETI@home is only application specific. The Extraterrestrial Intelligence (SETI) is a university of California project. It developed desktop grid system that has hundred and thousand of Pc's across Internet to process massive amount of astronomy data capturing through Arecibo telescope based at Puerto Rico every day.

Condor **[7]** system is developed by university of Wisconsin at Madison. Unique mechanism enable candor to effectively harness wasted CPU Power from idle desktop workstations. Uses submit their job to condor, condor places them into a queue, chooses when and where to run the jobs based upon a policy monitors is the progress and informs upon completion. It can handle both windows and Unix class resources in its resource pools. This does not have tread-programming model and does not support cross platform web services interface.

|  | Alchemi | Condor | SETI@home | Entropia | Nimble@ITCEcnoGrid |
|---|---|---|---|---|---|
| Architecture | Hierarchical | Hierarchical | Centralized | Centralized | Hierarchical with centralized database |
| Web Services Interface for cross-platform Integration | Yes | No | No | No | Yes |
| Implementation Technologies | C#, Web services & .NET Framework | C | C++, Win 32 | C++, Win 32 | C#, Web services & .NET Framework 3.5 |
| Thread programming Model | Yes | No | No | No | Yes |
| Level of Integration of application, Prog. & runtime Environment | Low (General Purpose) | Low (General Purpose) | High (Single Purpose Single Application Environment) | Low (General Purpose) | High (Single Purpose Single Application Environment) |
| Inter thread Communication | No | No | No | No | Yes |

Table 1: Comparison of Nimble@ITCEcnoGrid and some related Enterprise Grid Systems[8]

The present research grid developed i.e Nimble@ITCEcnoGrid have Hierarchical Architecture with centralized database. The implementation technologies used in the development are C# & .NET Framework 3.5. It has thread programming model with inter thread communication. It is used to find the PI value up to 120 decimal digits after decimal & First order model equations of weather forecasting.

## 3. H/W & S/W REQUIREMENTS/INSTALLATION OF THE PREREQUISITES

The H/W and S/W used for the implementation of "Nimble@ITCEcnoGrid" are:
The H/W & S/W requirements for the test case is

1) 1 GB RAM
2) 1.8 GHz Core to Dual Processors
3) Network Interface Card
4) 4 pot Network Switch
5) SQL Server 2008 R2
6) Visual Studio 2008 SP1 or VS 2010 Beta 2
7) Windows Server 2008 R2

Next Step is to install Installation of Server 2008 [9], SQL 2008 [10] and then VS 2010 [11]

## 4. METHODLOGY

Nimble@ITCEcnoGrid is a web based Interface, which can be accessed & controlled by the Administrator. There are Nimble@ITCEcnoGrid Managers at different locations i.e. different managers for different locations. There can be multiple managers at one location This is because, the failure of one manager at some location will not affect the work. Data after calculation is send to database of Nimble@ITCEcnoGrid Web Interface, which will display the results. There are number of Executors connected to the Sub manager which are in turn connected to Manager. Executors after doing the required task will submit the result to the Sub Manager, which in turn send result to Manager. Manager will further send result to the Database of Nimble@ITCEcnoGrid Website. More the number of executors, less time the manager will take to produce result. The executors can work on real IP's or LAN or VPN because if it get disconnected in between for some reason, after reconnection the IP address will remain same so that cost of contributing the free cycles can be added economically in the database. The benefit of using the above Model is that, it provides the reliability which was not there in Alchemi.Net as instead of one Manager whose failure can affect the Whole Grid there can be number of Managers at one location and failure of any one or more manager will not stop the work. Figure 1 shows the Model of the Nimble@ITCEcnoGrid.

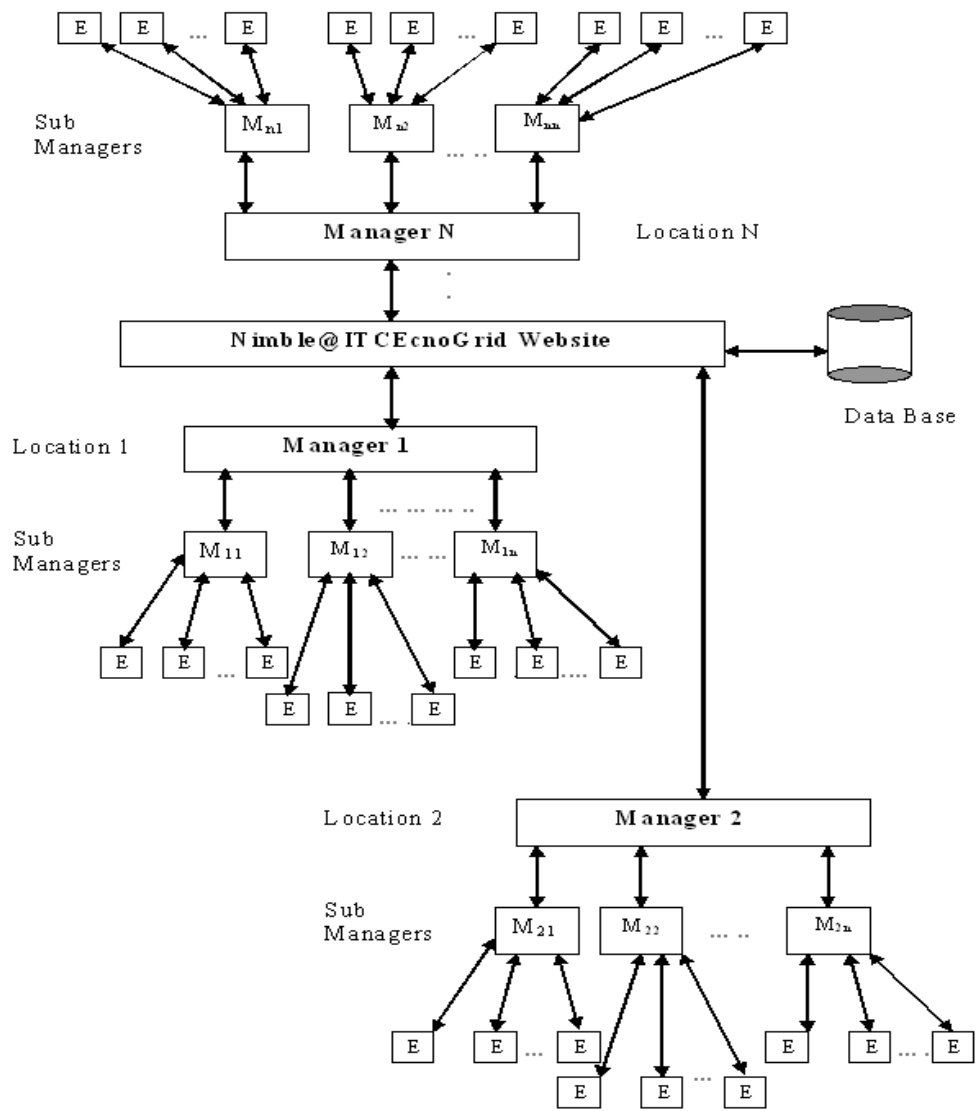

Fig 1: Model of Nimble@ITCEcnoGrid

International Journal of Grid Computing & Applications (IJGCA) Vol.2, No.4, December 2011

## 4.1. Installation of Nimble@ITCEcnoGrid Manager, Sub Manager & Executor

### 4.1.1 Installation of Nimble@ITCEcnoGrid Manager

The Manager manages the execution of grid applications and provides services associated with managing thread execution. It is deployed as an executable. It then send the calculated to the Nimble@ITCEcnoGrid Website. The Screenshot of the Nimble@ITCEcnoGrid Manager is shown in fig2.

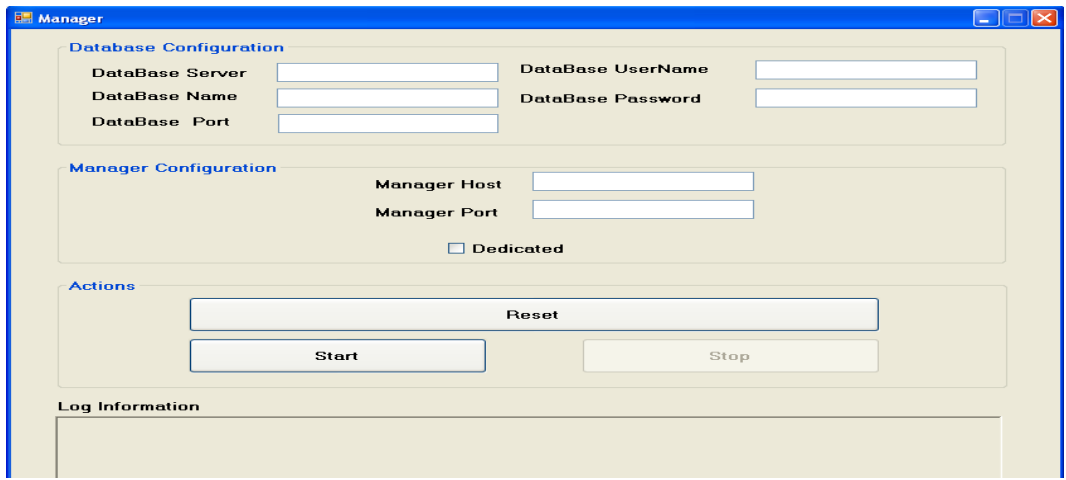

Fig 2: Nimble@ITCEcnoGrid Manager

### 4.1.2 Installation of Nimble@ITCEcnoGrid Sub Manager

The Sub Manager executes & Manages individual grid threads and provides services associated with executing threads. It is deployed as an executable. The Screenshot of the Nimble@ITCEcnoGrid Sub Manager is shown in fig 3.

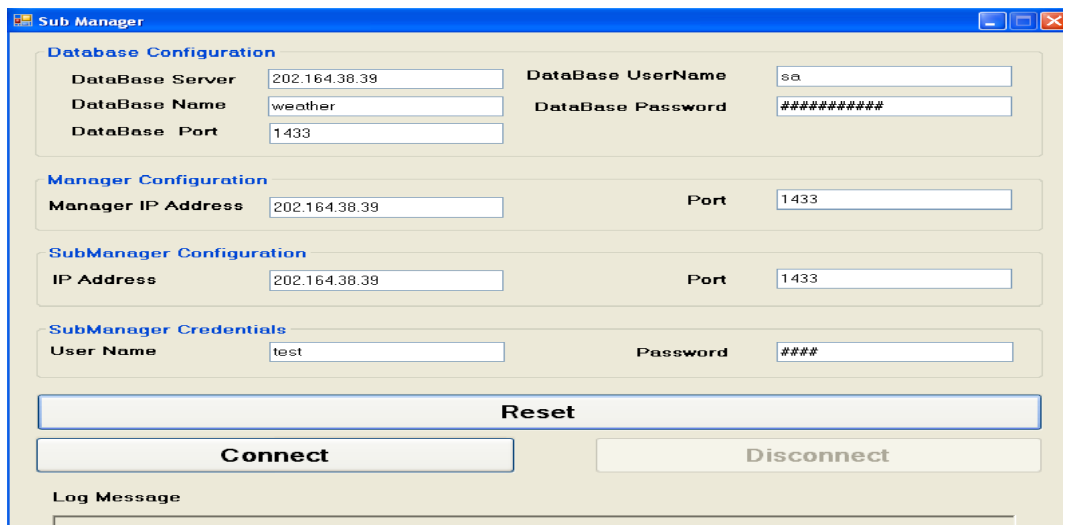

Fig 3: Nimble@ITCEcnoGrid Sub Manager

### 4.1.3 Installation of Nimble@ITCEcnoGrid Executor

The Executor executes individual grid threads and provides services associated with executing threads. It is deployed as an executable. The Screenshot of the Nimble@ITCEcnoGrid Executor is shown in fig 5.

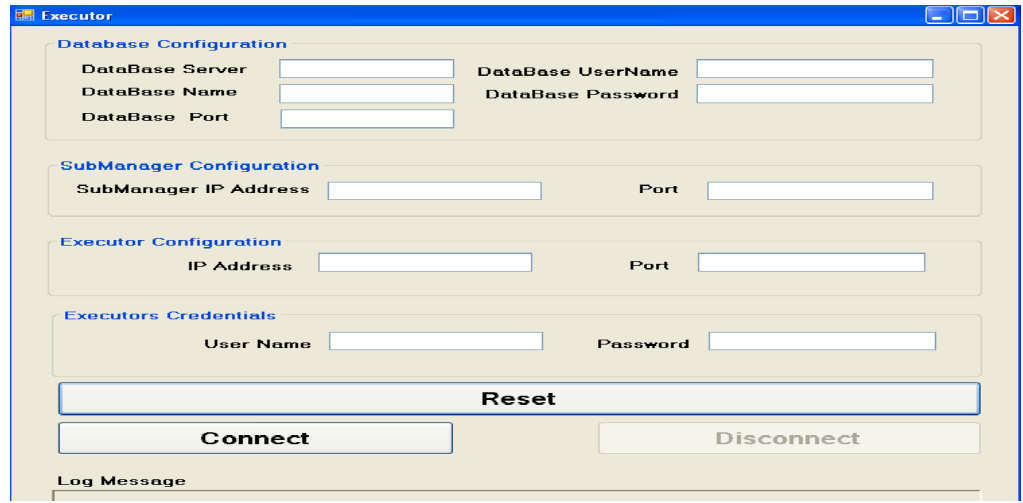

Fig 4: Nimble@ITCEcnoGrid Executor

## 4.2 PI value Calculation up to 120 digits after decimal places

Pi is one of the most important numbers in mathematics. It is defined as the ratio of a circle's circumference to its diameter, but it crops up in all sorts of places in mathematics. It is an infinitely long non-recurring decimal number. There are many ways to do this. Some methods converge rapidly but are complicated to implement. Some are simple to implement but converge very slowly. I've chosen a method that is fairly simple and converges reasonably fast. It is based on the following formula:

$$pi / 4 = 4 * \tan^{-1}(1 / 5) - \tan^{-1}(1 / 239)$$

This is Machin's formula[12] . **$\tan^{-1}()$** is the Inverse Tangent function, and I use the Maclaurin series [13] to calculate it:

$$\tan^{-1}(z) = z - z^3 / 3 + z^5 / 5 - z^7 / 7 + ...$$

By including sufficiently many terms of this series, we can achieve any desired accuracy. To get 1,000,000 decimal places accuracy for pi, we need about 715,000 terms of the $\tan^{-1}(1/5)$ series and about 210,000 terms of the $\tan^{-1}(1/239)$ series. I have used the program to calculate 120 decimal places accuracy.

## 4.3 Weather Forecasting Model Equations [14]

Weather prediction is a problem of predicting the future evolution of the atmosphere for minutes to days to perhaps 2 weeks ahead. It begins with observations of the initial state (and their uncertainties) and analyses into global fields, then use of a model of the atmosphere to predict all of the future evolution of the turbulence and eddies for as long as is possible. Because the atmosphere is a chaotic fluid, small initial uncertainties or model errors grow rapidly in time and make deterministic prediction impossible beyond about 2 weeks.
According to the Kevin E Trenberth *National Center for Atmospheric Research Boulder, Colorado USA[12]*,the Governing laws e.g.  for the Atmosphere ,the following equations are used as the first order model for the weather forecasting numerical calculations.

- *Momentum equations:*

d**V**/dt = -∇αp -2Ω^**V** –g**k** +F +Dm
where  α=1/ρ (ρ is density), p is pressure, Ω is rotation rate of the Earth, g is acceleration due to gravity (including effects of rotation), **k** is a unit vertical vector, F is friction and Dm is vertical diffusion of momentum

- *Thermodynamic equation:*

dT/dt = Q/cp + (RT/p)ω + DH
where cp is the specific heat at constant pressure, R is the gas constant, ω is vertical velocity, DH is the vertical diffusion of heat and Q = Qrad + Qcon   is internal heating from radiation and condensation/evaporation;

- *Continuity equations, e.g. for moisture (similar for other tracers):*

dq/dt = E – C + Dq
where E is the evaporation, C is the condensation and Dq is the vertical diffusion of moisture
Mainly the above defined equations gives the weather forecasting basic equations. In my thesis the above equations are used to compute the value of the Momentum, Thermodynamic & Continuity equation.

## 5. RESULTS

Based on the equations described above, the results calculated by the Nimble@ITCEcnoGrid are as follows:

### 5.1 Result of Equations using Internet as Medium

| No. of Executors | PI Value (Up to 120 digits after decimal) | Momentum | Continuity | Thermodynamics |
|---|---|---|---|---|
| 1 | 3.14159265358979323846264338327950288419716939937510582097494459230781640686208998628013482534211706798214808651328230664 7<br>Time: 19 ms | Value:11.2854765770015<br>Time : 932 ms | Value:1.009<br>Time :2 ms | Value:257745.930080227<br>Time : 70 ms |
| 2 | 3.14159265358979323846264338327950288419716939937510582097494459230781640686208998628013482534211706798214808651328230664 7<br>Time: 5 ms | Value:11.2854765770015<br>Time : 28ms | Value:1.009<br>Time :8 ms | Value:257745.930080227<br>Time : 27 ms |
| 3 | 3.14159265358979323846264338327950288419716939937510582097494459230781640686208998628013482 | Value:11.2854765770015<br>Time : 210 ms | Value:1.009<br>Time :14 ms | Value:257745.930080227<br>Time : 828 ms |
| | 4211706798214808651328 2306647 Time: 75 ms | | | |
| 4 | 3.14159265358979323846 26 43383279502884197169399 37510582097494459230781 64068620899862801348253 4211706798214808651328 2306647 Time: 5 ms | Value:11.285476 5770015 Time : 4 ms | Value:1.009 Time :3ms | Value:257745.93008022 7 Time : 992ms |

Table 2: Result of Equations for Weather Forecasting using Internet as medium

## 5.2 Result of Equations using LAN as Medium

| No. of Executo rs | Momentum | Continuity | Thermodynamics | PI Value (Up to 120 digits after decimal) |
|---|---|---|---|---|
| 1 | Value:11.2854765770 015 Time : 741 ms | Value:1.009 Time :11 ms | Value:257745.9300 80227 Time : 129 ms | 3.1415926535897932384626 4338327950288841971693993 7510582097494459 23 0781640686208998628 01348 2534211706798214808651 32 82306647 Time: 25 ms |
| 2 | Value:11.2854765770 015 Time : 1578ms | Value:1.009 Time :15ms | Value:257745.9300 80227 Time : 11 ms | 3.1415926535897932384626 4338327950288841971693993 7510582097494459 23 0781640686208998628 01348 2534211706798214808651 32 82306647 Time: 23 ms |
| 3 | Value:11.2854765770 015 Time : 165ms | Value:1.009 Time :7 ms | Value:257745.9300 80227 Time : 2 ms | 3.1415926535897932384626 4338327950288841971693993 7510582097494459 23 0781640686208998628 01348 2534211706798214808651 32 82306647 Time: 9 ms |
| 4 | Value:11.2854765770 015 Time : 40 ms | Value:1.009 Time :3ms | Value:257745.9300 80227 Time : 4ms | 3.1415926535897932384626 4338327950288841971693993 7510582097494459 23 0781640686208998628 01348 2534211706798214808651 32 82306647 Time: 11 ms |

Table 3: Result of Equations for Weather Forecasting using LAN as medium

## 5.3 Economic Based Policy

The following figure shows the amount earned by the executor by giving its free cycles of the computer. The rate of dedicated computers is .02 Rs per pulse. The table 3 shows that IP no 202.164.38.39 has worked for 160 seconds and has earned Rs 3.20 .

| Sr. No. | Manager Id | Manager Host | Start DateTime | Stop DateTime | Work In Sec | Rate/Sec | Cost |
|---|---|---|---|---|---|---|---|
| 1 | 1 | 192.168.1.9 | 7/17/2011 11:35:00AM | 7/17/2011 11:35:00AM | 0 | 2.00 | 0.00 |
| 2 | 2 | CurrentTime | 7/17/2011 11:35:00AM | 7/17/2011 11:35:00AM | 0 | 2.00 | 0.00 |
| 3 | 3 | 192.168.1.99 | 7/17/2011 11:35:00AM | 7/17/2011 11:35:00AM | 0 | 2.00 | 0.00 |
| 4 | 4 | 192.168.1.101 | 7/17/2011 11:35:00AM | 7/17/2011 11:35:00AM | 0 | 2.00 | 0.00 |
| 5 | 5 | 192.168.1.5 | 7/17/2011 11:35:00AM | 7/17/2011 11:35:00AM | 0 | 2.00 | 0.00 |
| 6 | 6 | 192.168.1.55 | 7/17/2011 11:35:00AM | 7/17/2011 11:35:00AM | 0 | 2.00 | 0.00 |
| 7 | 7 | 192.168.1.100 | 7/17/2011 11:35:00AM | 7/17/2011 11:35:00AM | 0 | 2.00 | 0.00 |
| 8 | 8 | 192.168.1.4 | 7/17/2011 11:35:00AM | 7/17/2011 11:35:00AM | 0 | 2.00 | 0.00 |
| 9 | 9 | 192.168.1.14 | 7/17/2011 11:35:00AM | 7/17/2011 11:35:00AM | 0 | 2.00 | 0.00 |
| 10 | 10 | 202.164.38.32 | 7/17/2011 11:35:00AM | 7/17/2011 11:35:00AM | 0 | 2.00 | 0.00 |
| 11 | 11 | 202.164.38.39 | 7/16/2011 11:11:11AM | 7/16/2011 5:11:11PM | 160 | 2.00 | 3.20 |
| 12 | 12 | 192.168.14.254 | 7/16/2011 11:11:11AM | 7/16/2011 5:11:11PM | 0 | 2.00 | 0.00 |

Total Amount : 3.20

Table 4  Economic Based Policy

## 6. CONCLUSION

A Computational InterGrid (Nimble@ITCEcnoGrid) having distributed resources based on .NET framework is developed to calculate value of 'PI' & the first order weather forecasting model. Nimble@ITCEcnoGrid is having inter-thread connectivity which does not exist in any kind of Windows/Linux based Grid. Economic based recourse allocation policy is developed which can be refined or renewed according to the specific policy.

The results clearly indicates that the Nimble@ITCEcnoGrid framework model can be used as a good future scope for solving the complex real time problems requiring high performance computers like weather forecasting & climate assessment in a research mode.

## 7. LIMITATIONS

As the real time problems need a time bound & higher precision calculations, Nimble@ITCEcnoGrid may be used in an Research Domain and not independently in operational domain until fast gigabits or even higher bandwidth network becomes available.

## 9. FUTURE SCOPE

The future scope of the research is to implement the Nimble@ITCEcnoGrid Model in the real time problems like Climate Prediction, Earthquake Prediction etc with high speed networks. As the future of Grid computing is Cloud, so it can be associated with the cloud computing to solve the real time problems.

## REFERENCES

[1]   Moore, Gordon E. (1965). *"Cramming more components onto integrated circuits" Electronics Magazine*". Retrieved on 2006-11-11.

[2]   Ian Foster and Carl Kesselman (editors), "*The Grid: Blueprint for a Future Computing Infrastructure*", Morgan Kaufmann Publishers, USA, 1999.

[3]   Ian Foster, Carl Kesselman, and S. Tuecke, "*The Anatomy of the Grid: Enabling Scalable Virtual Organizations*", *International Journal of Supercomputer Applications*, 15(3), Sage Publications, 2001, USA.

[4]   David Anderson, Jeff Cobb, Eric Korpela, Matt Lebofsky, Dan Werthimer, "*SETI@home: An Experiment in Public-Resource Computing*", *Communications of the ACM,* Vol. 45 No. 11, ACM Press, USA, November 2002.

[5]   Akshay Luther, Rajkumar Buyya, Rajiv Ranjan, and Srikumar Venugopal, "*Alchemi: A .NET-Based Enterprise Grid Computing System*", Proceedings of the 6th International Conference on Internet Computing (ICOMP'05), June 27-30, 2005, Las Vegas, USA.

[6]   Andrew Chien, Brad calder, Stephen Elbert and Karan Bhatia,"*Entropia: Architecture and Performance of an Enterprise Desktop Grid System*", Journal of parallel and Distributed Coputing, Volume 63, Issue 5, Academics Press, USA, May 2003.

[7]   Rajiv Ranjan, Aaron Harwood, Rajkumar Buyya-"*SLA-Based Coordinated Superscheduling Scheme and Performance for Computational Grids*" -In Proceedings of the 8th IEEE International Conference on Cluster Computing (Cluster 2006), IEEE Computer Society Press, September 27 - 30, 2006, Barcelona, Spain.

[8]   James Broberg · Srikumar Venugopal · Rajkumar Buyya-"*Market-oriented Grids and Utility Computing:The State-of-the-art and Future Directions*" Received: 18 September 2007 / Accepted: 12 December 2007 / Published online: 28 December 2007 © Springer Science + Business Media B.V. 2007

[9]   www.petri.co.il/how-to-install-windows-server-2008-step-by-step.htm

[10]   http://msdn.microsoft.com/en-us/library/ms143219.aspx

[11]   stuff.seans.com/2009/visual-studio-2010-install-screenshots

[12]   http://mathworld.wolfram.com/MachinsFormula.html

[13]   http://mathworld.wolfram.com/MaclaurinSeries.html

[14]   Climate and weather forecasting: Issues and prospects for prediction of climate on multiple time scales byKevin E Trenberth *National Center for Atmospheric Research,Boulder, Colorado USA, International Symposium on Forecasting, June 24-27 2007 (Some slides borrowed from others: esp Bill Collins)*

## Authors

**Er. Vijay Dhir[1]** is working as a Assistant Professor in Information Technology at Sant Baba Bhag Singh Institute of Engineering & Technology, Padhiana, Distt: Jalandhar, Punjab, India.. He has a work experience of 11 years. He has published 8 International and 7 National Papers at various Journals/Conferences.

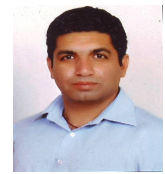

**Dr. Rattan K Datta[2]** is working as a C.E.O & Director, MERIT, New Delhi,India. He retired as the Adviser ,Department of Science & Technology, Government of India. He did his Ph.D from IIT (Delhi) on "Monsoon Dynamics & Modelling".He has guided number of students in Ph.D, M.Tech & M.Phil. He has published about 200 research papers in various International & National Journals/Conferences.

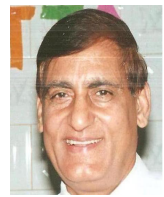

**Dr. Maitreyee Dutta3** is working as a Associate Professor at NITTTR,Chandigarh,India. She has a working experience of 5 years in Research and 11 years in Teaching. She has guided 40 M.Tech students. She has published 12 papers in various International journals & conferences.

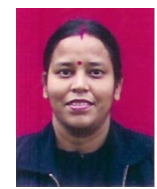